\documentclass[a4paper,11pt]{article}
\pdfoutput=1

\usepackage{jinstpub} 
\usepackage{lineno}

\title{\boldmath The active muon shield in the SHiP experiment}

\author{

A.~Akmete$^{43}$, 
A.~Alexandrov$^{12}$,
A.~Anokhina$^{35}$,
S.~Aoki$^{16}$,
E.~Atkin$^{34}$,
N.~Azorskiy$^{25}$,
J.J.~Back$^{49}$,
A.~Bagulya$^{28}$,
A.~Baranov$^{36}$,
G.J.~Barker$^{49}$
% ...

\vspace*{1cm}

{\footnotesize \it

% ...
$ ^{12}$Sezione INFN di Napoli, Napoli, Italy\\
% ...
$ ^{16}$Kobe University, Kobe, Japan\\
% ...
$ ^{25}$Joint Institute of Nuclear Research (JINR), Dubna, Russia\\
% ...
$ ^{28}$P.N.~Lebedev Physical Institute (LPI), Moscow, Russia\\
% ...
$ ^{34}$National Research Nuclear University (MEPhI), Moscow, Russia\\
$ ^{35}$Skobeltsyn Institute of Nuclear Physics of Moscow State University (SINP MSU), Moscow, Russia\\
$ ^{36}$Yandex School of Data Analysis, Moscow, Russia\\
% ...
$ ^{43}$Middle East Technical University (METU), Ankara, Turkey\\
% ...
$ ^{49}$University of Warwick, Warwick, United Kingdom\\
% ...
}
}

\emailAdd{Hans.Dijkstra@cern.ch}

\abstract{The SHiP experiment is designed to search for very weakly interacting particles beyond the Standard Model which are produced in a 400 GeV/c proton beam dump at the CERN SPS. An essential task for the experiment is to keep the Standard Model background level to less than 0.1 event after $2\times 10^{20}$ protons on target. 
In the beam dump, around $10^{11}$ muons will be produced per second.
The muon rate in the spectrometer has to be reduced by at least four orders of magnitude to avoid muon-induced combinatorial  background.
A novel active muon shield is used to magnetically deflect the muons out of the acceptance of the spectrometer. This paper describes the basic principle of such a 
shield, its optimization and its performance.}


\collaboration{\includegraphics[height=17mm]{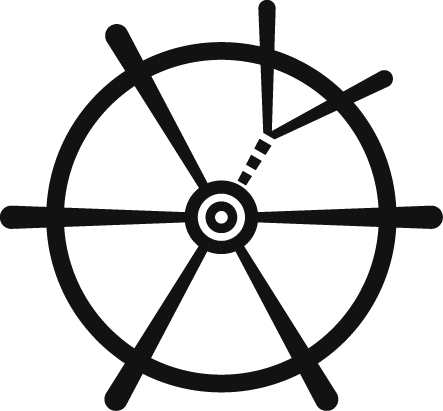}\\[6pt]
   The SHiP collaboration}

\begin{document}
\maketitle
\flushbottom

\section{Introduction}
\label{sec:intro}
The SHiP experiment \cite{TP} is a new general purpose fixed target facility proposed at the 
CERN SPS accelerator to search for new physics in the largely unexplored domain of very weakly interacting particles with masses below $\mathcal{O}(10)~\mathrm{GeV/c^2}$ and c$\tau$ of kilometers~\cite{PhysicsCase}.
The facility is designed to maximize the production of charm and beauty mesons, while providing the cleanest possible environment. 
The 400 GeV/c proton beam extracted from the SPS will be dumped into a molybdenum-tungsten target, 
with the aim of accumulating $2\times 10^{20}$ protons on target (POT) during five years of operation. 
The charm production at SHiP exceeds that of any existing or presently planned facility. 

For radiation protection, a 5 m thick iron shield is placed behind the target to absorb hadrons. The SPS will deliver $4\times 10^{13}$ POT
during a one second long spill, which produces on the order of $10^{11}$~muons/s mainly from the decay of $\pi$, $\rm{K}$, $\rho$, $\omega$ and charmed mesons. 
These muons would give rise to a serious background for many hidden particle searches, and hence their flux has to be reduced by at least four orders of magnitude. 
The SHiP experiment will use of a series of magnets to deflect the muons out of the acceptance of the spectrometer. 
Any hidden sector particle originating from charm or beauty meson decay will be produced with a significant transverse momentum with respect to the beam axis\footnote{In the SHiP coordinate system the z-axis is along the beam line and the y-axis is pointing upward.}. In order to maximize acceptance, the detector therefore should be placed as close as possible to the target, and hence the muon shield
should be as short as possible. Behind the muon shield 
a dedicated detector will probe a variety of models with long lived exotic particles with masses below $\mathcal O(10)~\rm{GeV/c^2}$.
The detector is designed to fully reconstruct 
the exclusive decays of hidden sector particles and to reject the background down to below 
0.1 events in the sample of $2\times 10^{20}$ POT.

The detector is based on a 
50~m long decay volume, housing a $5\times10~\rm m^2$ 
spectrometer magnet which is sandwiched between tracker stations. 
To suppress the background from neutrinos 
interacting in the fiducial volume, the decay volume is maintained under a vacuum since at atmospheric pressure $\sim 10^5$ neutrino interactions are expected
for $2\times 10^{20}$ POT. 
The spectrometer is designed to accurately reconstruct the decaying particle's decay vertex, its mass and 
its impact parameter.
Behind the last tracking station, a dedicated 
 timing detector with resolution less than 100 ps measures the coincidence of the 
decay products, which allows the rejection of combinatorial backgrounds.
The timing detector is followed by a set of calorimeters 
and by muon 
chambers to provide identification of electrons, photons, muons and charged hadrons. 
The decay volume is surrounded by background taggers to detect the products of neutrino and muon 
inelastic scattering in the surrounding structures, which may produce long lived 
Standard Model $\rm{V}^0$ particles, such as $\rm{K}_{\rm{L}}$, that have a similar decay topology to the expected signals.  

The experimental facility is also ideally suited 
for studying interactions of the least known tau neutrinos~\cite{DONUT}.
It will therefore host in between the muon shield and the decay volume an emulsion target with a maximum acceptable charged particle flux of $10^3$/mm$^2$ for an exposure based on the emulsion cloud  concept developed in OPERA~\cite{Opera} and a muon spectrometer. 

In the simulation, fixed target collisions of protons are generated by PYTHIA8~\cite{Pythia8}, neutrino interactions by GENIE~\cite{Genie} and inelastic muon interactions by 
PYTHIA6~\cite{Pythia6}. The heavy flavor production in cascade interactions is also taken into account~\cite{Cascade}. 
The SHiP detector response is simulated in the GEANT4~\cite{Geant4} framework. All the simulation is 
performed within the FairRoot~\cite{FAIRROOT} framework.

\begin{figure}[htbp]
\centering
\includegraphics[width=1.\textwidth,clip]{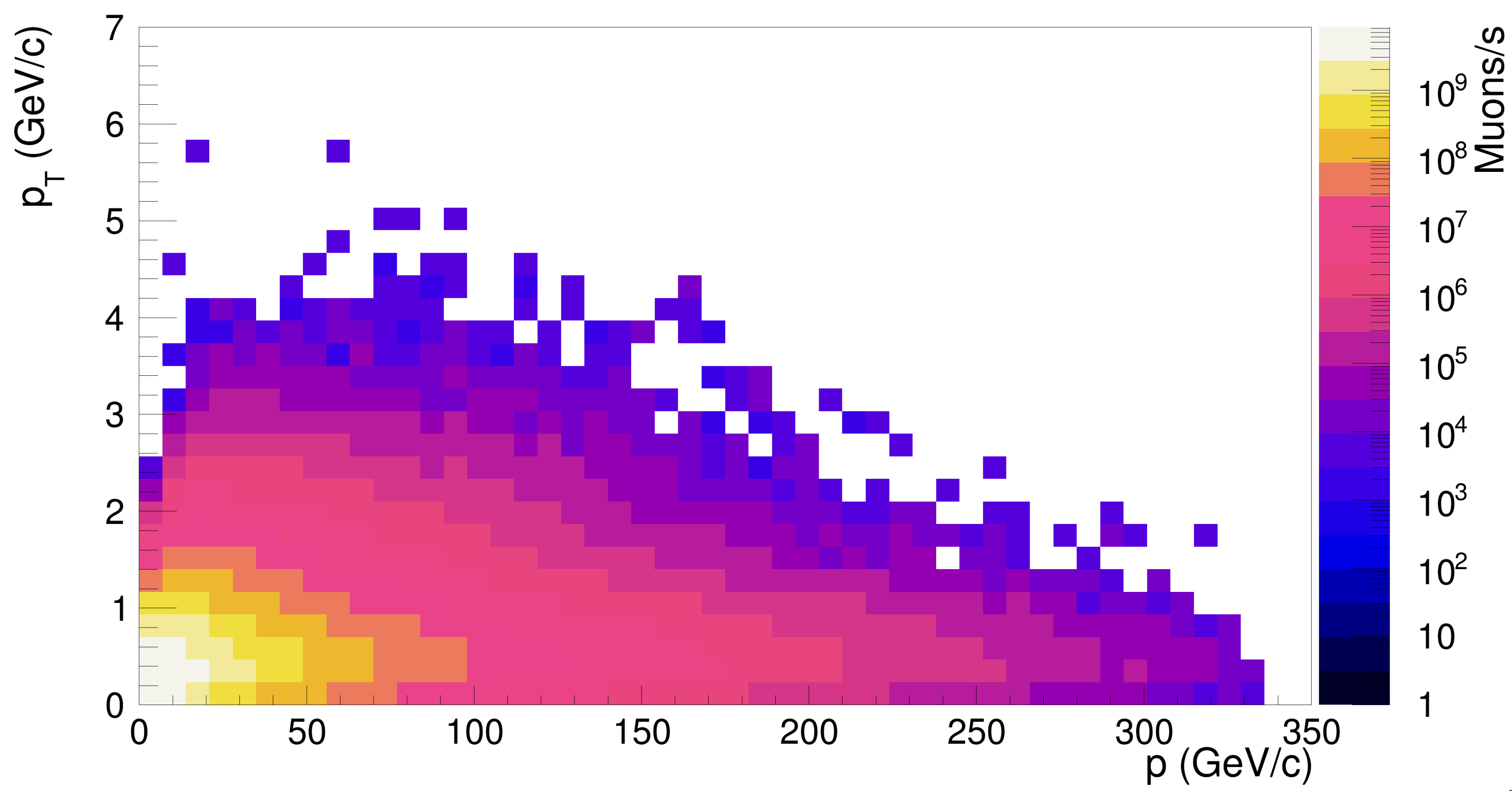}
\caption{Transverse momentum versus momentum distribution of muons, as generated by Pythia~\cite{Pythia8, Pythia6}.}
\label{fig:muonFlux_initial}
\end{figure}

While muons can easily be deflected out of the acceptance of the spectrometer by
a magnetic field, the problem is the large spread in the phase space of muons, as shown in figure~\ref{fig:muonFlux_initial}, with the consequence that the return field of the magnets tend to bend muons back towards the spectrometer.
The DONUT~\cite{DONUT} experiment employed a combination of magnetic and passive shielding
to clear muons from their emulsion target, which was located 36 m downstream
of the target. 
SHiP will accumulate three orders of magnitude more protons on target, and hence requires a much larger
reduction in muon flux for both the emulsion and the hidden sector particles search.
To achieve this SHiP will employ the novel magnetic shielding concept described in this paper.

The basic magnetic configuration that addresses these issues is presented in the next section.
Section~\ref{sec:simul} describes the optimization procedure with a simplified Monte Carlo tool to obtain a layout 
which reduces the muon flux sufficiently for a minimum size of the system.
The optimization is checked by implementing the obtained layout in the GEANT4-based SHiP software package, and its
performance is presented in section~\ref{sec:performance}. The last section discusses possible improvements
and gives an outlook for obtaining the final design of the SHiP muon shield.

\section{The basic principle of magnet based muon shielding}
\label{sec:principle}
\begin{figure}[htbp]
\centering
\includegraphics[width=0.6\textwidth,clip=]{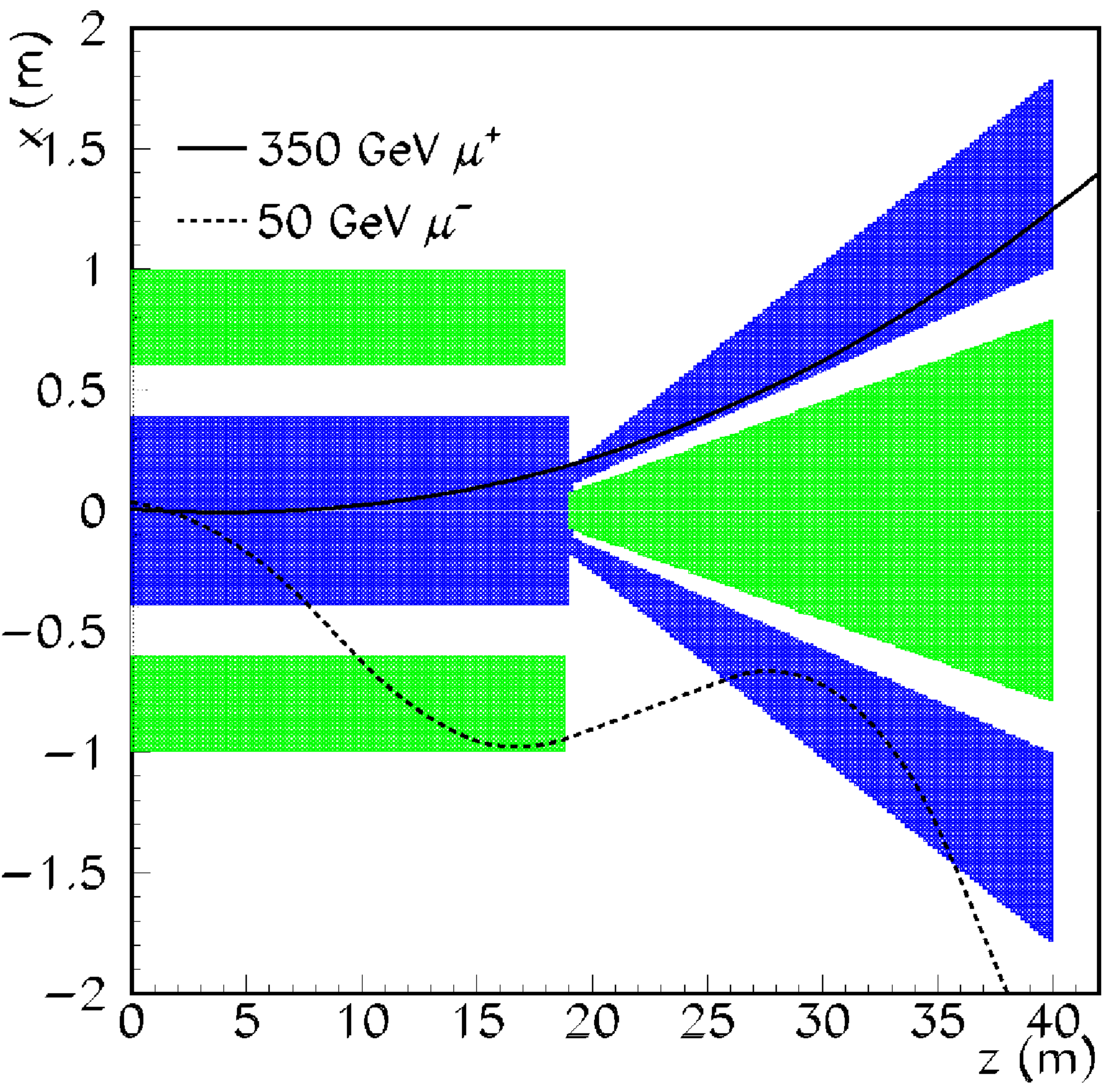}
\caption{Cross section at y=0 shows the principle of magnetic shielding. 
The magnetic field is along the y-axis, and its polarity is 
indicated by the blue/green color of the iron poles of the magnets.
The trajectories of a 350 GeV/c muon and 50 GeV/c muon are shown with a full and dashed line, respectively.}
\label{fig:peda}
\end{figure}

The principle of the magnet based shielding is shown in 2-D in figure~\ref{fig:peda}.
The first part of the shield should be long enough, i.e.
providing sufficient $\int Bdl$, to separate both muon polarities to either side of the z-axis. For a 350 GeV/c muon, taking into account the $p_T$
distribution at its production point, this requires a $\sim 18$ m long magnet
with a field\footnote{When using a high permeability material such as Grain Oriented steel a 1.8 T 
flux density could be obtained while the coils would dissipate low enough power so that they could be cooled with air~\cite{Vicky}.} in the iron of 1.8 T.
 
Lower momentum muons, which traverse the return field of this magnet, will be bent back in the direction of the spectrometer as is shown
in figure~\ref{fig:peda} for a 50 GeV/c muon.
To shield the spectrometer from these muons, an additional magnet
is added with opposite polarity field close to the z-axis, and hence
the lower momentum muons will be swept out again. At the start of
this second magnet, the two field polarities should be as close in x as
possible . A magnet design study~\cite{Vicky} shows that an air gap as small
as 2 cm between the two field polarities can be used without distorting the 1.8 T field in the iron.
\section{Simulations and optimization}
\label{sec:simul}
To test the basic principle of the muon shielding concept described in section~\ref{sec:principle} and optimize its layout, a fast program was written which
traces the muons through the magnets.
Each magnet is described by the following seven parameters: length and, at either end, widths, heights, and air-gap widths between field and return field.
Muons are traced through a setup with a step size of 5 cm, and
only the following physics processes are taken into account:
bending in a 1.8 T magnetic field, specific energy loss
($dE/dx$) and 
Gaussian smeared multiple Coulomb scattering.

The aim is to design the shortest possible shield which allows no muons to reach the spectrometer, while
at the same time being affordable and hence using as little iron as possible. These requirements are combined
into a single penalty function which is then minimized. The penalty function is given by 
\begin{equation}
\frac{ W\times (1+\Sigma\chi_{\mu^+})}{A} 
\end{equation}
where:
\begin{itemize}
\item All muons are assigned a positive charge, and the magnetic fields are oriented to bend the muons towards positive x at the last tracking station.
The transverse (x,y) position of muons is obtained at the last tracking station, which has a sensitive area of $5\times 10\rm ~m^2$,
by tracing their trajectory 64 m downstream of the shield.
Their position at the last tracking station is converted to a smooth function
to allow the optimization to sense that it moves all $\mu^+$ out of the acceptance of the last tracking station.
Hence, with $|y|<5$ m, their x-position is converted into
$\chi_{\mu^+} =\sqrt{(5.6-(x+3.))/5.6}$ for $-3<x<2.6$ m, else  $\chi_{\mu^+} =0$;
\item $A=1.-L/100$: Approximately the relative acceptance of a new long lived particle in a benchmark scenario, where $L$ is the length of the shield in m;
\item $W$: Weight of the shield in tonnes of iron;
\item The sum is taken over all muons.
\end{itemize}
The optimization uses Minuit~\cite{Minuit} to minimize penalty function  (3.1) by
varying the size of the magnets. 
Only muons with $p(\mu)>1$ GeV/c at the last tracking station are considered as background muons, since only larger momentum tracks will be used in the searches. Magnet transverse sizes are only allowed to vary by quanta of 1 cm.
Magnet length sizes are only allowed to vary by quanta of 10 cm.
To assure that a muon which traverses the same geometry
returns the same contribution to the penalty function the tracing
of each muon is always started with a random number seeded by its momentum.
For simplicity, the optimization is made with seven magnets. Future work will try to vary the number of magnets. The first two magnets after the beam dump
are also required for radiation protection and their lengths are fixed to 1.4 and 
3.4 m respectively, their gaps are fixed to
2 cm and they have the same height at either end of each magnet.

Muons have been generated with Pythia~\cite{Pythia8, Pythia6}, corresponding to $\sim 10^{10}$ POT, and the Minuit SIMPLEX function varies 41 parameters to minimize the penalty function.
After convergence, none of the muons traverse the last tracking station
for a total shield length of 34 m. Its layout and performance is presented in the next section.

\section{Active muon shield performance}
\label{sec:performance}
\begin{figure}[htbp]
	\centering
	\includegraphics[width=0.6\textwidth,clip=]{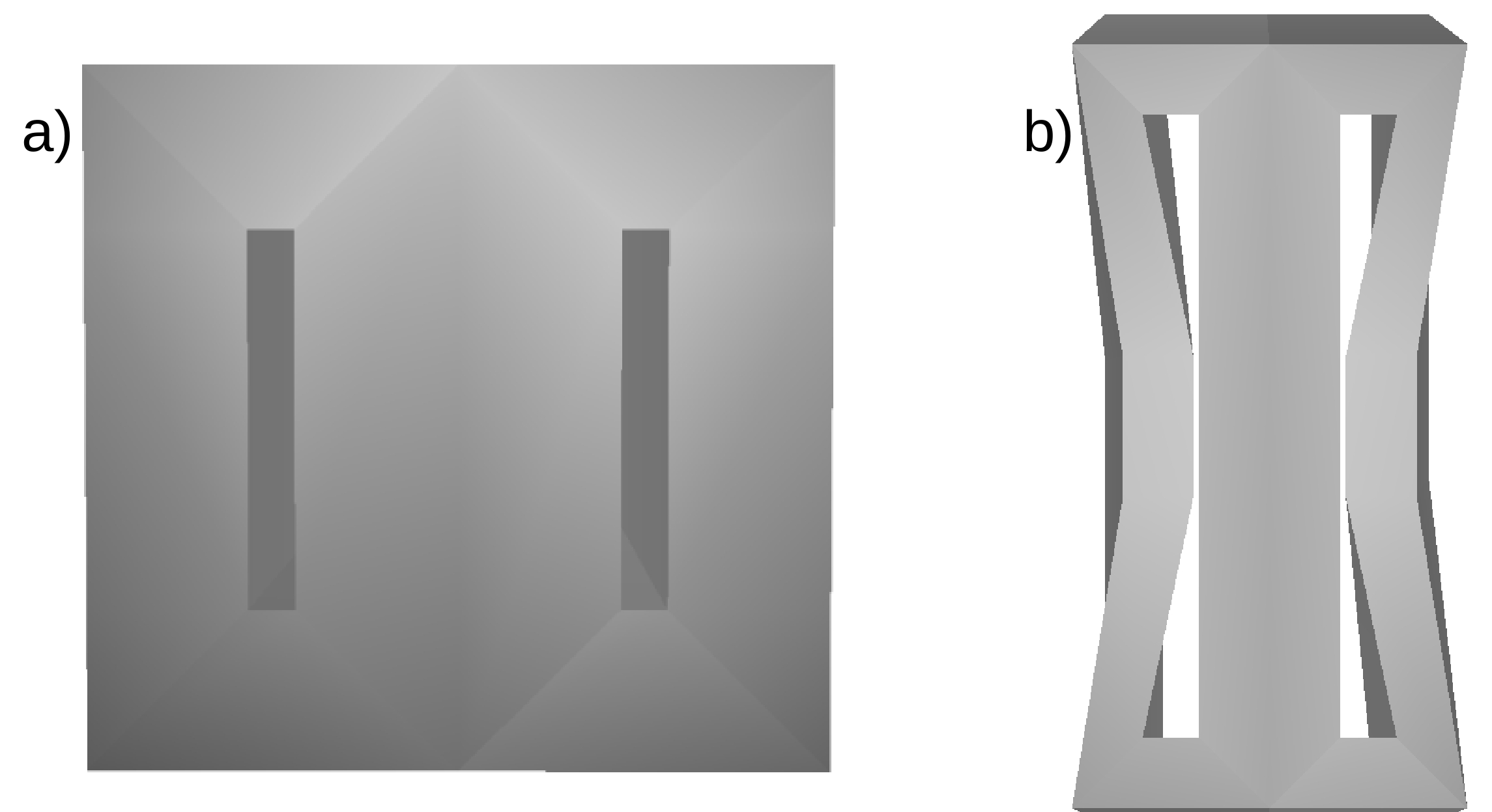}
	\caption{
	A view in the x-y plane of the two magnet configurations. 
In configuration a) the air-gap is large enough anywhere along y to place a coil. Configuration b) has an air-gap close to 2 cm at y=0, and hence the program
will create large enough air-gaps at the top and bottom of the magnet to 
accommodate the coils.
}
	\label{fig:magnets}
\end{figure}

The SHiP experiment simulations are performed in the software package FairShip~\cite{TP}. FairShip is mostly a collection of libraries and scripts based on the FairRoot framework~\cite{FAIRROOT}, which is fully based on the ROOT software framework~\citep{root}. In the framework, users can construct their detectors, and perform simulation and analysis without any code dependence on a specific Monte Carlo simulation software. 
FairRoot uses the ROOT geometry package TGeo to build, browse, track and visualize a detector geometry. 
The muon shield can be made of two different magnet types, that are depicted
in figure~\ref{fig:magnets}.  
When there is enough air-gap between the iron yokes
with the opposite field direction to place a coil
a magnet of type a) in figure~\ref{fig:magnets} is used. If the optimization requires
an air-gap which is too small to accommodate a coil, a magnet
of type  b) in figure~\ref{fig:magnets}, which has place for coils at the top
and bottom, is used.

\begin{figure}[htbp]
	\centering
	\includegraphics[width=\textwidth,clip=]{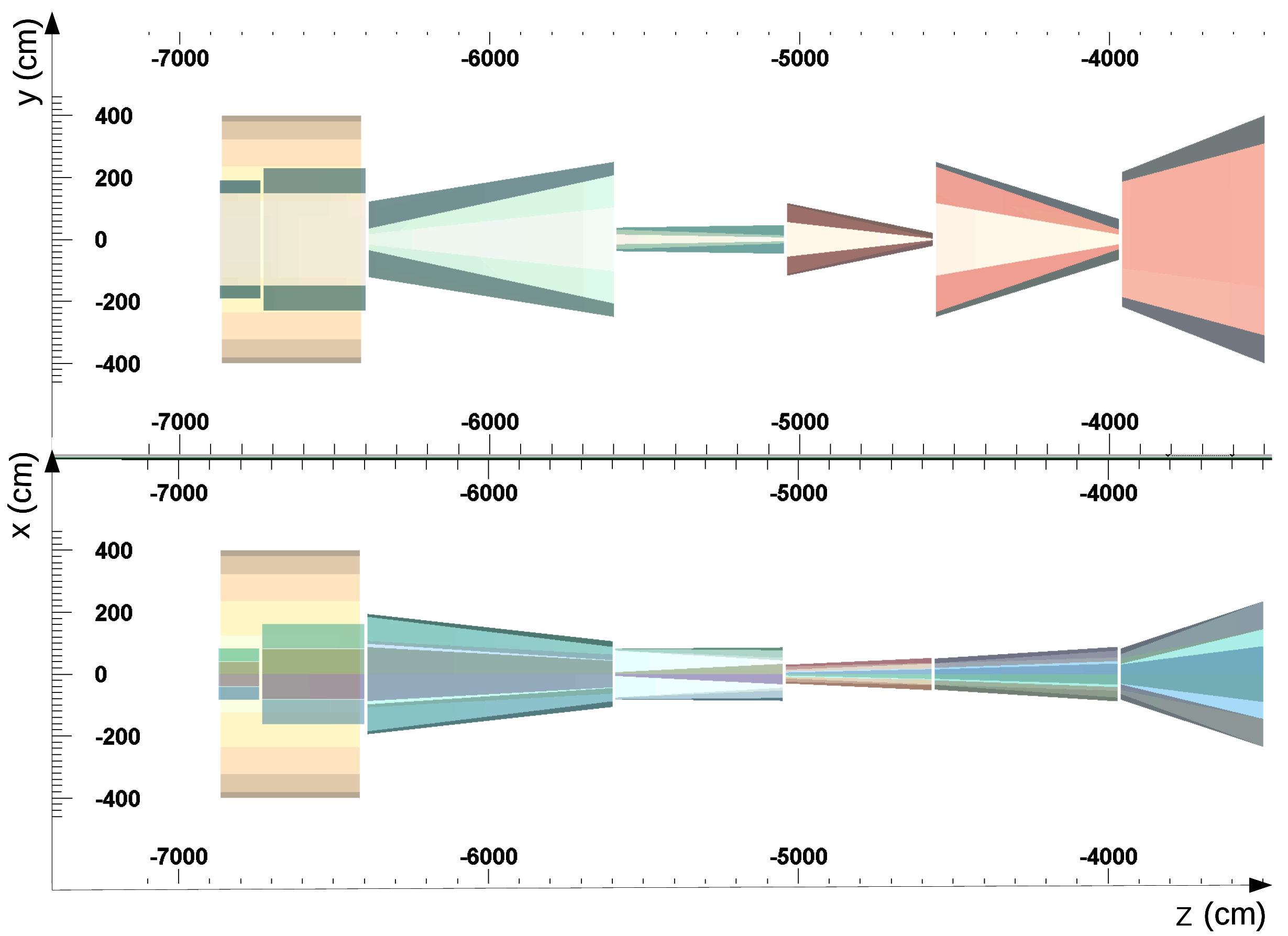}
	\caption{Geometric view of the optimized muon shield, showing at the top, the 
z-y plane view, and at the bottom, the z-x plane view. SHiP defines the origin 
of the coordinate system to be in the center of the decay vessel. 
Color shading is used to enhance the contrast between different magnetic field orientations.
}
	\label{fig:Shield}
\end{figure}

In addition, the following boundary conditions are imposed to assure a realistic implementation of the magnets:
\begin{itemize}
\item For every cross section in the xy-plane, the width of the iron is
large enough everywhere to sustain the same magnetic flux;
\item Minimum air-gap between opposite field directions is 2 cm;
\item Minimum air-gap for accommodating coils in the plane perpendicular to the beam direction is 20 cm;
\item Minimum air-gap between magnets along the beam direction is 10 cm;
\item Mitred joints between volumes with horizontal and vertical
 	fields are imposed to lower their magnetic reluctance~\cite{Vicky}.
\end{itemize}
The geometry optimization results in a shield layout shown in figure~\ref{fig:Shield},
weighing 1845 tonnes, which does not include the supports.

To test the background with the prototype shielding in FairShip, only target,
 	muon shield, tracking stations and spectrometer magnet were included in the simulation. 
 	The last tracking station is located 64 m
 	downstream of the last shielding magnet.

The simulation is first run with all the material of the shield in place, but
with the magnetic field switched off. This results in $\sim 3.1\times 10^9$ muons
which traverse the last tracking station during one SPS spill of $4\times 10^{13}$ POT. 
Switching on the magnetic field 
reduces this rate to $\sim 3.0\times 10^5$ muons/s, of which $\sim 6.5\times 10^4$ have 
momenta larger than 3 GeV/c. These muon rates are sufficiently low
to cause insignificant background levels in the experiment.

Examining the muons which do reach the last tracking station reveals that they are
mainly due to two effects which are not included in the optimization procedure, 
namely catastrophic energy loss~\cite{cata} and very large scattering angles.

\section{Discussion and outlook}
A magnetic deflecting system for muons 
is presented, which provides the required rate reduction of muons for the SHiP 
experiment and fulfills the emulsion requirements described in the
introduction.
The design is optimized to reduce muon background in the spectrometer for the shortest possible shield
and the lowest iron usage.
The optimized geometry of the shield is 34 m long and its weigth is 1845 tonnes not including its supports. 
As coming from FairShip simulation, given that the muons which do traverse the system are due to
physics processes which have not been included in the optimization program,
the shield will be further optimized using all physics processes as 
implemented in GEANT4~\cite{Geant4}. This will increase the required CPU time by several
orders of magnitude.
In addition, it is found that a large flux of electrons and photons is 
produced by the muons traversing the magnets.
Their momentum is typically below a GeV/c, and hence they do not constitute
a source of background directly. However, the SHiP decay vessel is surrounded by 
detectors to tag any charged particle entering the vessel. Moreover, the emulsion target upstream is also sensitive to low energy electrons that may spoil its reconstruction capability. Hence, the flux 
of these particles has to also be controlled to ensure an effective
use of these surround taggers. A future optimization will include the
flux of electrons and photons through the taggers and emulsion detectors in the penalty function.

\end{document}